\documentclass[aps,twocolumn,showpacs,groupedaddress,nofootinbib]{revtex4}
\usepackage{epsfig}
\usepackage{bm}

\newcommand{\Pythia}{\textsc{pythia}}
\newcommand{\Jetset}{\textsc{jetset}}

\begin{document}

\preprint{02-02}

\title{ Deeply inelastic
pions in the exclusive reaction $p(e,e'\pi^{+})n$ above the resonance region}
\author{Murat M. Kaskulov}
\email{Murat.Kaskulov@theo.physik.uni-giessen.de}
\author{Kai Gallmeister}
\author{Ulrich Mosel}
\affiliation{Institut f\"ur Theoretische Physik, Universit\"at Giessen,
             D-35392 Giessen, Germany}
\date{\today}

\begin{abstract}
A model for the $p(e,e'\pi^+)n$  reaction which combines an improved 
treatment of gauge invariant meson--exchange currents and  hard
deep--inelastic  scattering (DIS) of virtual 
photons off nucleons is proposed. It is shown that DIS
dominates and explains the transverse response at  moderate and high
photon virtualities $Q^2$ whereas the longitudinal response is  dominated by
hadronic degrees of freedom and the pion electromagnetic form factor. 
This leads to a  combined description of the
longitudinal and transverse components of the cross  section in a wide range
of photon virtuality $Q^2$ and momentum transfer  to the target $t$ and solves 
the longstanding problem of the observed large transverse cross sections. The
latter are shown to be sensitive to the intrinsic transverse momentum
distribution of partons.
\end{abstract}
\maketitle

At Jefferson Laboratory (JLAB) the exclusive
reaction $p(e,e'\pi^+)n$  has been investigated for a range of photon
virtualities up to $Q^2\simeq 5$~GeV$^2$ at an invariant mass of the $\pi^+n$
system around the onset of deep--inelastic regime,
$W\simeq 2$~GeV~\cite{Horn:2006tm,Tadevosyan:2007yd,Horn:2007ug}.
A separation of the cross section into the transverse $\sigma_{\rm T}$
and longitudinal $\sigma_{\rm L}$ components has been performed.
The longitudinal cross section $\sigma_{\rm L}$ is well understood
in terms of the pion quasi--elastic knockout
mechanism~\cite{Neudatchin:2004pu}
because of the pion pole at low $-t$ . This makes it possible to
study the charge form factor of the pion at momentum transfer much bigger
than in the scattering of pions from atomic electrons~\cite{Sullivan:1970yq}.
On the other hand, the 
$\sigma_{\rm T}$ is
predicted to be suppressed by $\sim 1/Q^2$ with respect to $\sigma_{\rm L}$
for sufficiently high $Q^2\gg \Lambda_{\rm QCD}^2$~\cite{Collins:1996fb}.

However, the data from the $\pi$--$CT$ experiment~\cite{Horn:2007ug} show
that $\sigma_{\rm T}$ is large at JLAB energies.
At $Q^2 = 3.91$~GeV$^2$ $\sigma_{\rm T}$ is by about a factor of two
larger than $\sigma_{\rm L}$
and at $Q^2=2.15$~GeV$^2$ it has same size as $\sigma_{\rm L}$ in agreement
with previous JLAB measurements~\cite{Horn:2006tm}. Theoretically, the model
of Ref.~\cite{Vanderhaeghen:1997ts}, which is generally considered to be a
guideline for the experimental analysis and extraction of the pion form
factor, underestimates $\sigma_{\rm T}$ at $Q^2=2.15$~GeV$^2$ and at
$Q^2=3.91$~GeV$^2$  by about one order of magnitude~\cite{Horn:2007ug}.
Previous measurements at values of $Q^2=1.6~(2.45)$~GeV$^2$~\cite{Horn:2006tm} 
show a similar problem in the understanding of $\sigma_{\rm T}$. 
Even at smaller JLAB~\cite{Tadevosyan:2007yd} and 
much higher Cornell~\cite{Bebek:1977pe}
values of $Q^2$ there is a disagreement between model
calculations  based on the hadron--exchange scenario and experimental data;
see Ref.~\cite{Faessler:2007bc} for a possible interpretation and
references therein.

\begin{figure*}[t]
\begin{center}
\includegraphics[clip=true,width=1.8\columnwidth,angle=0.]
{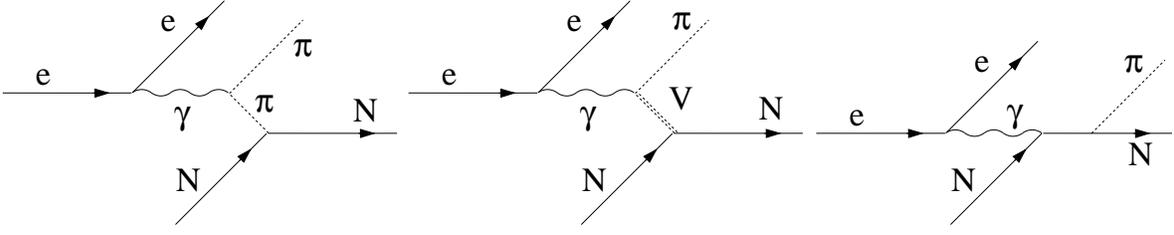}
\caption{\label{Figure1} 
\small The diagrams describing 
       the hadronic part of the $\pi^+$-- electroproduction amplitude at high
       energies. See text for
       the details. \vspace{-0.5cm} }
\end{center}
\end{figure*}

In this work we first generalize the treatment of
Ref.~\cite{Vanderhaeghen:1997ts}
for the longitudinal contribution. We then propose a resolution of
the $\sigma_{\rm T}$ problem. The idea followed here is to complement the soft
hadron--like interaction types shown in Figure~\ref{Figure1} 
which dominate in photoproduction and low $Q^2$ electroproduction 
by {direct} hard interaction of virtual photons with partons 
followed by the hadronization process into $\pi^+n$ channel,  
to form the $\pi^+$-- electroproduction
framework. As we shall show, then the large $\sigma_{\rm T}$ in the reaction 
$p(e,e'\pi^+)n$ can be readily explained and both $\sigma_{\rm L}$ and 
$\sigma_{\rm T}$ can be described from low up to high values of $Q^2$.

The exclusive reaction 
\begin{equation}
e(P_e) + N(p) \to e'(P_e') + \pi(k') + N(p')
\end{equation}
with unpolarized electrons is described by four structure functions
$\sigma_{\rm T }$, $\sigma_{\rm L }$, $\sigma_{\rm LT}$ and
$\sigma_{\rm TT}$~\cite{Hand:1963bb}. After the integration over the
azimuthal angle between the leptonic and hadronic scattering planes only
$\sigma_{\rm T }$ and $\sigma_{\rm L }$ remain and the  differential cross
section takes the form
\begin{eqnarray}
{d\sigma_{e} }/{dQ^2d\nu dt} &=&
\frac{\pi \, \Phi}{E_e(E_e-\nu)}
\left[
           {d \sigma_{\rm T }}/{dt}
+ \varepsilon {d \sigma_{\rm L }}/{dt} \right],
\end{eqnarray}
where $\varepsilon$ is the virtual photon polarization. The definition of
the virtual photon flux $\Phi$ follows the convention of Ref.~\cite{Hand:1963bb}.
The subscripts  {\textsc t} and {\textsc l} denote the projections of the
$(\gamma^*,\pi)$ amplitude onto the basis vectors $\epsilon_{\mu}^{\lambda}$
of the circular polarization of the virtual photon
quantized along its three momentum $\vec{q}$: {\textsc t}-- transverse
($\lambda=\pm1$) and {\textsc l}-- longitudinal ($\lambda=0$) polarizations.

At first we consider the soft hadron--exchange part of the
$\pi^+$--electroproduction  amplitude. In Figure~\ref{Figure1} the Feynman
diagrams describing the high energy $\pi^+$ electroproduction in the
hadron--exchange  approach are shown. It has been well known for a long
time that the $\pi$--pole amplitude, first diagram in
Figure~\ref{Figure1}, gives the  dominant contribution to the longitudinal
response $\sigma_{\rm L}$. The $\pi$--pole amplitude by itself is not
gauge invariant and charge conservation requires an addition of the
electric part of the $s$--channel nucleon Born term (third diagram in
Figure~\ref{Figure1})~\cite{Jones:1979aa,Vanderhaeghen:1997ts}. When
considering  realistic vertex functions, which include  form factors,
current conservation is violated and one has to restore the gauge
invariance of the model~\cite{Naus:1989em}.
A simple solution to this problem is to choose all of the electromagnetic 
form factors to be  the same~\cite{Vanderhaeghen:1997ts}. However, it is 
experimentally known that these form factors are not the same and have 
different scaling behavior $F_{\gamma\pi\pi} \sim 1/Q^2$ for the pion 
form factor and $F_1^p\sim1/Q^4$ for the proton Dirac form factor.

In the following we use the Regge pole model of Ref.~\cite{Vanderhaeghen:1997ts}
which is based on the same set of Born diagrams but concerning the 
electromagnetic form factors we employ a prescription proposed in 
Refs.~\cite{Koch:2001ii,Gross:1987bu}  where an arbitrary form factor 
$F(Q^2)$ can be accommodated by the following replacement of the currents
\begin{equation}
\label{PT}
{\it \Gamma}^\mu \rightarrow {{\it \Gamma}'}^\mu(Q^2)= {\it \Gamma}^\mu
+[F(Q^2)-1] \,
\mathcal{P}^{\mu \nu}_{\perp} {\it \Gamma}_\nu\,,
\end{equation}
where $\mathcal{P}^{\mu \nu}_{\perp}=g^{\mu \nu}-q^{\mu} q^{\nu}/q^{2}$
stands for the projector into the $3$-dimensional transverse subspace.
This procedure guarantees that the resulting current ${{\it \Gamma}'}^\mu$ 
obeys the same Ward--Takahashi identities as ${\it \Gamma}^\mu$. Thus, and 
as long as gauge invariance is implemented for  real photons,
one can use the experimentally determined form factors in the 
$\pi$--pole $J^{\mu}_{\pi}$ and $s$--channel nucleon Born $J^{\mu}_{s}$
currents and still retain gauge invariance for arbitrary $Q^2$.

Making use of Eq.~(\ref{PT}) the $\gamma\pi\pi$  and  $\gamma NN$
vertex functions are given by
\begin{eqnarray}
\label{GammaPi}
{\it \Gamma}_{\gamma\pi\pi}^{\mu}
= e\,(k+k')^{\mu}+e\,[F_{\gamma\pi\pi}(Q^2)-1] \,\mathcal{P}^{\mu
  \nu}_{\perp}(k+k')_{\nu}, \\
\label{GammaPi2}
{\it \Gamma}_{\gamma NN}^{\mu} = e\,\gamma^{\mu}+e\,[F_{1}^N(Q^2)-1] \,\mathcal{P}^{\mu \nu}_{\perp}
\gamma_{\nu}, \hspace{1cm}
\end{eqnarray}
where the four momentum vectors of pions are $k$ (incoming) and $k'$
(outgoing). In Eq.~(\ref{GammaPi}) we have - as usual - assumed that the  
half--off--shell form factor $F_{\gamma \pi \pi}(Q^2,t)$ depends only on $Q^2$.
Using Eqs.~(\ref{GammaPi}) and~(\ref{GammaPi2})
the gauge invariant hadronic current $J^{\mu}$ 
describing the reaction $p(\gamma^*,\pi^+)n$ is
constructed as a sum 
$J^{\mu} = J^{\mu}_{\pi}+J^{\mu}_{s}$.

At high energies the exchange of high--spin and high--mass particles
has to be taken into account. To account for these states we replace 
in $J^{\mu}_{\pi}$ the $\pi$--Feynman propagator  by the
Regge propagator~\cite{Vanderhaeghen:1997ts}. Furthermore, since the 
$s$--channel Born term can generate the pion pole itself~\cite{Jones:1979aa}, 
we factor out the pion propagator in the sum $J^{\mu}_{\pi}+J^{\mu}_{s}$
following Ref.~\cite{Vanderhaeghen:1997ts} and reggeize it according to 
the above prescription. The hadronic current which satisfies the current 
conservation, {\it i.e.} $q_{\mu} J^{\mu} =0$, takes the form
\begin{eqnarray}
\label{PipoleFin}
{-iJ^{\mu}} =
{\sqrt{2} g_{\pi NN}} \, \bar{u}_{s'}(p')\gamma_5 
\left[
F_{\gamma\pi\pi}(Q^2) \frac{ (k+k')^{\mu}}{t-m_{\pi}^2+i0^+} 
\right.
 \nonumber \\
+   F_{1}^p(Q^2) \frac{k'_{\sigma}\gamma^{\sigma}
 \gamma^{\mu}}{W^2-M_p^2+i0^+}
\nonumber \\
+ \left.
[F_{\gamma\pi\pi}(Q^2)-F_{1}^p(Q^2)] \frac{(k-k')^{\mu}}{Q^2} 
\right] u_s(p) 
\nonumber \\
\times  [t-m_{\pi}^2+i0^+] \left(\frac{W}{W_0}\right)^{2 \alpha_{\pi}(t)}
\hspace{-0.5cm}
\frac{\pi \alpha'_{\pi}}{\sin(\pi\alpha_{\pi}(t))}
\frac{e^{-i\pi\alpha_{\pi}(t)}}{\Gamma(1+\alpha_{\pi}(t))},
\end{eqnarray}
where 
\begin{equation}
\alpha_{\pi}(t)=\alpha^0_{\pi}+\alpha'_{\pi}t = 0.7 (t-m_{\pi}^2)
\end{equation}
is  the degenerate $\pi$--$b_1$--trajectory, $W_0=1$~GeV and the Gamma
function $\Gamma$ suppresses the singularities in the physical region ($t<0$).
In Eq.~(\ref{PipoleFin}) $g_{\pi NN}=13.4$ is the pseudoscalar $\pi N$ coupling constant,
$t=k^2$, $k=k'-q=p-p'$ and other notations are obvious.
It should be noted that in the current (\ref{PipoleFin})
the two  different form factors of the nucleon and the pion appear;  this is in
contrast to the work of \cite{Vanderhaeghen:1997ts} where these two
form factors $F_{\gamma \pi \pi}$ and $F_1^p$  were assumed to be identical in 
order to reach gauge invariance. 
For the pion charge form factor we use a monopole parameterization
\begin{equation}
F_{\gamma\pi\pi} (Q^2) =
{(1+Q^2/\Lambda_{\gamma\pi\pi}^2)^{-1}},
\end{equation} 
with the cut--off $\Lambda_{\gamma\pi\pi}$ as a fit parameter.
The 
Dirac form factor $F_1^p(Q^2)$ is described by a standard 
dipole form.

\begin{figure*}[t]
\begin{center}
\includegraphics[clip=true,width=1.86\columnwidth,angle=0.]{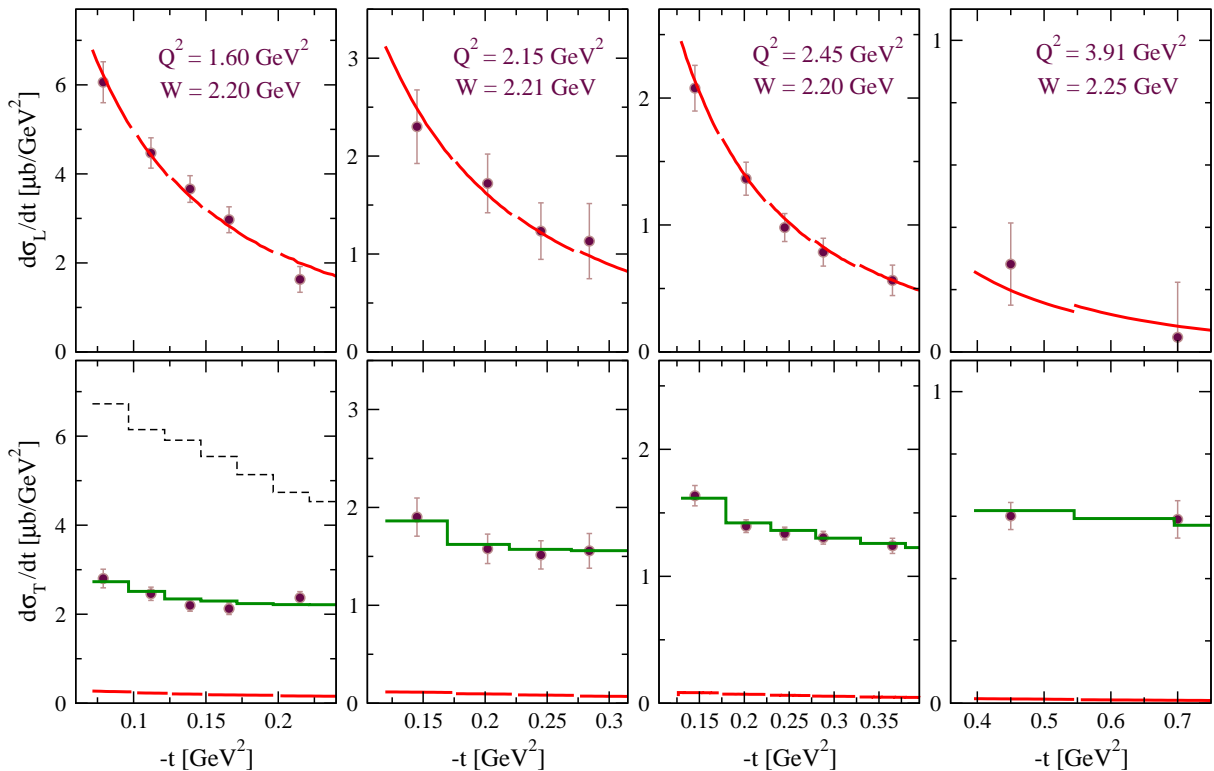}
\caption{\label{Figure3}
The longitudinal $d\sigma_{\rm L}/dt$~(top panels) and
transverse $d\sigma_{\rm T}/dt$ (bottom panels) differential cross sections
of the reaction $p(\gamma^*,\pi^+)n$ at average values of
$Q^2=1.60~(2.45)$~GeV$^2$~\cite{Horn:2006tm}
and $Q^2=2.15~(3.91)$~GeV$^2$~\cite{Horn:2007ug}.
The solid curves are the contribution of the hadron--exchange model
and the histograms are the contribution of the DIS pions.
The discontinuities in the curves result from
the different values of $Q^2$ and $W$ for the various $-t$ bins. 
The dashed histogram in the lower
left panel shows  the contribution of the DIS pions
for the average transverse momentum of partons $\sqrt{\langle k_{\rm t}^2\rangle}=0.4$~GeV. 
\vspace{-0.5cm}
}
\end{center}
\end{figure*}

The second diagram in Figure~\ref{Figure1} describes the exchange
of the $\rho$--meson Regge trajectory. The current
$J_{\rho}^{\mu}$ reads
\begin{eqnarray}
\label{rho}
-iJ_{\rho}^{\mu} &=& - i \sqrt{2}  \, G_{\rho NN} G_{\gamma\rho\pi}
F_{\gamma\rho\pi}(Q^2)
{\varepsilon^{\mu\nu\alpha\beta} q_{\nu}
 k_{\alpha}}
\nonumber \\
&& \times
\bar{u}_{s'}(p') \left[  (1 + \kappa_{\rho}) \gamma_{\beta}
- \frac{\kappa_{\rho}}{2M_p} (p+p')_{\beta} \right] u_s(p)  \nonumber \\
&& \times \left( \frac{W}{W_0} \right)^{2\alpha_{\rho}(t)-2} \hspace{-0.5cm}
\frac{\pi \alpha'_{\rho}}{\sin(\pi \alpha_{\rho}(t))}
\frac{e^{-i\pi \alpha_{\rho}(t)}}
    {\Gamma(\alpha_{\rho}(t))}.
\end{eqnarray}
The parameters needed for the proper description of the current $J_{\rho}^{\mu}$ 
are 
\begin{equation}
\alpha_{\rho}(t)= 0.55+0.8t
\end{equation}
as  the degenerate $\rho$--$a_2$--trajectory, 
$G_{\rho NN}=3.1$ is the vector and $\kappa_{\rho}=6.1$ is the tensor
$\rho N$ coupling constants. The ${\gamma\rho\pi}$ coupling constant
$G_{\gamma\rho\pi} = 0.728$ GeV$^{-1}$ has been deduced from the
decay width~\cite{PDG} 
\begin{equation}
\Gamma_{\rho^+ \to \pi^+ \gamma} \simeq 67.5\mbox{~keV}.
\end{equation}
For the $\gamma \rho \pi$ vertex form factor $F_{\gamma\rho\pi}$
we use the prediction of Ref.~\cite{Maris:2002mz}.

In Figure~\ref{Figure3} the results
for the $p(\gamma^*,\pi^+)n$ differential cross sections $d\sigma_{\rm
L}/dt$ (top panels) and $d\sigma_{\rm T}/dt$~(bottom panels) are compared
with the data from  JLAB $F\pi$--2~\cite{Horn:2006tm} and
$\pi$--$CT$~\cite{Horn:2007ug} experiments. The longitudinal cross section
$d\sigma_{\rm L}/dt$ is very well described by the hadron--exchange 
model (solid curves) with the cut--off in the pion form factor being fixed 
to the constant value  $\Lambda_{\gamma\pi\pi}^2=0.52~$GeV$^2$. 
The discontinuities in the curves result from
the different values of $Q^2$ and $W$ for the various $-t$ bins. 
The steep fall of $d\sigma_{\rm L}/dt$ away from forward
angles comes entirely from the rapidly decreasing $\pi$--pole amplitude.
The interference of this amplitude with the $s$--channel nucleon Born term is
minimized due to the presence of different form factors for both amplitudes.
The contribution of the natural parity $\rho$--exchange is negligible
in $\sigma_{\rm L}$ and $\sigma_{\rm T}$.

This comparison with data shows that $F_{\gamma\pi\pi}$ can indeed be reliably
extracted from the longitudinal data.

Again, the model strongly underestimates $d\sigma_{\rm T}/dt$,
for example, at $Q^2=1.6$~GeV$^2$ by a factor of 10 and at
$Q^2=3.91$~GeV$^2$ by a factor of 30.
This is also seen in the model of Ref.~\cite{Vanderhaeghen:1997ts}, although
somewhat less pronounced~\cite{Horn:2006tm,Horn:2007ug}.

The solution to this problem is still missing. One might describe this 
transverse strength in the language of perturbative QCD by considering higher 
twist corrections to a GPD based handbag diagram. However, such a calculation 
does not exist and it is not clear if a higher--twist expansion converges in 
the kinematical regime considered here. Our solution of this problem, 
therefore, is to model such effects. We start from 
the observation that
the second term  of Eq. (\ref{PipoleFin})
contributes very little to both the longitudinal and transverse cross sections.
Here only the nucleon Born term is taken into
account to conserve the charge of the system.
However, at the  invariant masses reached in the experiment ($W
\approx 2.2$ GeV)  nucleon resonances can contribute to the $1\pi$ channel.
Similar to the replacement above of the pion propagator by a Regge propagator
that takes higher meson excitations into account we now complement
the $s$--channel nucleon Born term with direct hard interaction of virtual photons with
partons (DIS)  since  DIS involves all possible transitions of the
 nucleon from its ground state to any excited state~\cite{Close:2001ha}. 
Note, that our suggestion concerning the partonic  
contribution follows the qualitative arguments 
in~\cite{Nachtmann:1976be}
where it has been shown that the typical exclusive photoproduction 
mechanisms involving a
peripheral quark--antiquark pair in the proton wave function, 
the $t$--channel
meson--exchange processes considered above, should be unimportant in the
transverse response already around
$Q^2 \gtrsim 1$~GeV$^2$ and play no role in the true deep inelastic region. This
we have already seen in Figure~\ref{Figure3}.

The total transverse DIS cross section reads
\begin{equation}
\label{DISnucleon}
\sigma_{\rm T}^{\rm DIS} 
= \frac{4 \pi^2 \alpha}{
  1-{x}} \frac{F_1^p({x},Q^2)}{\nu M_p}
 =\frac{4 \pi^2 \alpha}{
  1-{x}} \frac{F_2^p({x},Q^2)}{Q^4}
\frac{Q^2+ 4M_p^2 x^2}{1+\mathcal{R}({x},Q^2)}, 
\end{equation}
where $\alpha \simeq \frac{1}{137}$ and ${x} = \frac{Q^2}{2\nu M_p }$.
In the following we assume 
that a partonic description of 
deep--inelastic structure functions $F_{1(2)}^p(x,Q^2)$
works well
not only in the Bjorken limit where 
$\mathcal{R}\equiv\sigma_{\rm L}^{\rm DIS}
/\sigma_{\rm  T}^{\rm DIS}$ tends to zero but 
is valid down to values of $Q^2$ 
considered in Figure~\ref{Figure3}.

\begin{figure}[t]
\begin{center}
\includegraphics[clip=true,width=0.95\columnwidth,angle=0.]
{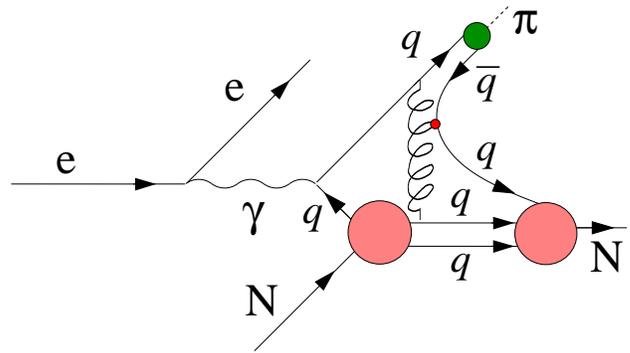}
\caption{\label{Figure1DIS} 
\small 
A schematic representation of the partonic part of the $\pi$--
electroproduction mechanism. The wavy line represents a color string. 
See text for the details.
\vspace{-0.8cm} }
\end{center}
\end{figure}

To determine the structure of events in DIS a model for the hadronization
process is needed. Furthermore, since at JLAB 
(Bjorken $x\gtrsim 0.3$) the antiquark content of the structure 
functions becomes negligible, we model the DIS by the 
$\gamma^* q \to q$ knockout reaction followed by hadronization through 
string fragmentation. In the present description of 
hadronization in DIS
we rely on the Lund model (LM)~\cite{Andersson:1983ia} 
as depicted in  Figure~\ref{Figure1DIS} where the 
$\gamma^* q \to q$ process followed by the fragmentation of an excited 
colored string (wavy curve connecting the quark lines\footnote{Not to be 
confused with the perturbative one gluon exchange.}) into two 
particles ($\pi N$) is shown. The LM predicts two jets 
for the $\pi^+n$ final state in the forward and backward directions.
As a realization of the LM in DIS we use the \Pythia{}/\Jetset{}
implementation~\cite{Sjostrand:2006za}. The LM involves parameters
which have been tuned in different fragmentation channels.
Our approach here is to modify as few parameters as 
possible compared
to the default set of values~\cite{Sjostrand:2006za}  
which  describes   the $\pi^+$ SIDIS spectra measured 
at JLAB~\cite{Avakian:2003pk} remarkably well. 
Since in \Pythia{} the average transverse momentum of partons
$\sqrt{\langle k_{\rm t}^2\rangle}$ cannot be fixed from first principles and 
since it affects the slope and magnitude of $d\sigma_{\rm T}/dt$ at 
forward angles we choose this as a free parameter. 
Therefore, one has to regard the average
$\sqrt{\langle k_{\rm t}^2\rangle}$ used here
as an effective parameter which is tuned to obtain an agreement with
data. However, for consistency we use the same value for 
$\sqrt{\langle k_{\rm t}^2\rangle}$ in all 
kinematic regimes together with the default \Jetset{} parameters.
As pointed out above we view the string fragmentation process as an 
effective model for higher--order twist effects, for example in GPD based 
handbag calculations. The success of our description may then be taken as 
an indication that the string fragmentation process described in  \Jetset{} 
works well down to the rather low invariant mass of about 2 GeV where 
the individual nucleon resonances tend to disappear.

The lower part of Figure~\ref{Figure3} shows that $d\sigma_{\rm T}/dt$
receives the dominant contribution from DIS fragmentation pions 
(solid histograms). 
In Eq.~(\ref{DISnucleon}) for $F_2^{p}$ we use the fit 
of Ref.~\cite{Abramowicz:1997ms} and 
for $\mathcal{R}$ the parameterization of 
Ref.~\cite{Tvaskis:2006tv} has been employed. 
In~\cite{Sjostrand:2006za} the value of 
$\sqrt{\langle k_{\rm t}^2\rangle}=1.2$~GeV has been used for all $Q^2$ bins; 
this value is close to the default  \Pythia{} value of 
$\sqrt{\langle k_{\rm t}^2\rangle}=1$~GeV and well within  the common range 
of transverse  momentum distributions~\cite{Close:1977mr}. As one can see in 
Figure~\ref{Figure3} (bottom panels) the absolute value and the $-t$ dependence
of $d\sigma_{\rm T}/dt$ are very well reproduced. 
A decrease of 
$\sqrt{\langle k_{\rm t}^2\rangle}$ increases the slope and magnitude 
of $d\sigma_{\rm T}/dt$ at forward  angles. 
In Figure~\ref{Figure3} this is shown 
for $\sqrt{\langle k_{\rm t}^2\rangle}=0.4$~GeV (dashed histogram).

We have also compared the model results with data from the
JLAB $F\pi$--1 experiment~\cite{Tadevosyan:2007yd}
at lower values of $Q^2$.
Also here we find that an addition of the DIS pions
describes the experimental data very well.
However, contrary to the situation at higher values of $Q^2$ where the 
hadronic part gives only a marginal
contribution to $\sigma_{\rm T}$, at low $Q^2$ 
the problem of double counting arises
when using both the DIS and the Regge contributions to the transverse cross
section. Following Ref.~\cite{Friberg:2000nx} 
this could be solved by turning off the leading order  DIS
contribution, as required by gauge
invariance for $\gamma^* q \to q$, when approaching the photon point 
where the Regge description
alone gives a good description of data~\cite{Vanderhaeghen:1997ts}. In the
calculation presented here 
the transverse part is solely generated by the DIS process 
(\ref{DISnucleon}) without any further modification. 

\begin{figure*}[t]
\begin{center}
\includegraphics[clip=true,width=1.6\columnwidth,angle=0.]{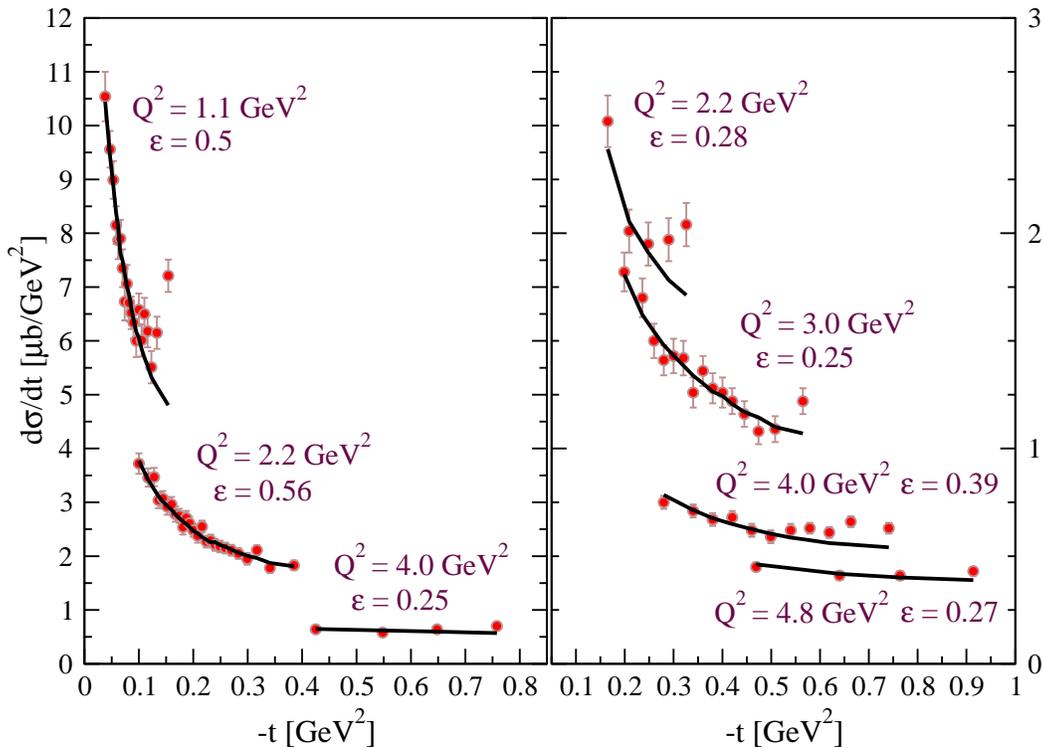}
\caption{\label{Figure4}
Differential cross section  
$d\sigma/dt = d\sigma_{\rm T}/dt + \varepsilon d\sigma_{\rm
  L}/dt$ of the reaction $p(\gamma^*,\pi^+)n$. 
The solid curves are the model predictions. The experimental data are 
from Ref.~\cite{Dutta}.
\vspace{-0.8cm}
}
\end{center}
\end{figure*}

In Figure \ref{Figure4} we confront the result of our calculations
(solid curves) with the new JLAB data~\cite{Dutta} 
for unseparated cross sections at average value of $W \simeq 2.2$~GeV. 
The data are very well described by the
present model in a measured range from $Q^2 \simeq 1$~GeV$^2$ 
up to $5$~GeV$^2$. 
Furthermore, assuming that 
the exclusive cross section behaves as 
$\sigma_{\rm T}^{\rm DIS}(Q^2)  \propto F_1^p(x,Q^2)$ in
Eq.~(\ref{DISnucleon})
\footnote{This behavior has been already noticed in~\cite{Bebek:1977pe} 
from a fit to data and is supported by the present model.}
and that the ratio $\mathcal R$ is small or nearly the same both for 
protons and neutrons
we predict then a smaller transverse cross section 
in the reaction $n(e,e'\pi^-)p$ off neutrons, 
{\it i.e.} 
\begin{equation}
\sigma^n_{\rm T}/\sigma^p_{\rm T} \simeq F_1^n/F_1^p \approx F_2^n/F_2^p < 1,
\end{equation} 
while  because of the $\pi$--pole dominance 
\begin{equation}
\sigma^n_{\rm L}/
\sigma^p_{\rm L}\simeq 1.
\end{equation} 
A preliminary analysis in Ref.~\cite{Horn:2006tm} has shown that the latter 
ratio 
is indeed consistent with unity and
$\sigma_{\rm L}/\sigma_{\rm T}$ must be larger for 
$\pi^{-}$ than for $\pi^+$.
This, together with the fact that 
$\sigma_{\rm L}$ is described very well also at the highest $Q^2$ by the 
Regge picture alone indicates that the DIS contribution to the exclusive 
longitudinal channel must be small.

In summary, in this work we have extended the earlier model of
Ref.~\cite{Vanderhaeghen:1997ts} such that the electromagnetic form factor of
the pion and the nucleon no longer have to be set equal in order to achieve
gauge invariance. In addition, we have proposed a
resolution of the $\sigma_{\rm T}$ problem in
the 
reaction $p(e,e'\pi^+)n$ above the resonance region.
A model which combines the gauge invariant 
hadron--exchange
currents and DIS of virtual photons off partons has been proposed.
The model with hadronic states as the active degrees of freedom describes the
longitudinal cross section $\sigma_{\rm L}$ very well
 and exhibits the dominance of the $\pi$--pole mechanism while 
$\sigma_{\rm T}$ is grossly underestimated.
We have shown that the description of $\sigma_{\rm T}$ at values of
$Q^2 > 1$~GeV$^2$
requires a proper inclusion of the hard scattering processes and
that $\gamma^* q \to q$ followed by the $\pi^+n$ fragmentation of the 
nucleon may
naturally explain the large transverse cross section observed at JLAB.
The model can be used for the extraction of
the pion form factor from high energy
pion electroproduction data with longitudinally polarized photons.
The sensitivity of the transverse cross section to the
transverse $\sqrt{\langle k_{\rm t}^2\rangle}$ of partons
can be used to 
reduce the theoretical uncertainties in the interpretation of
the color transparency signal observed at JLAB in the reaction $(e,e'\pi^+)$
off nuclei~\cite{:2007gqa,Kaskulov:2008ej}. 
Finally, we mention another 
$\sigma_{\rm L}/\sigma_{\rm T}$ puzzle in the reaction $p(e,e'
K^+)\Lambda(\Sigma)$~\cite{Niculescu:1998zj} which may apparently get a
similar solution~\cite{KM}.

\begin{acknowledgments}
We are grateful to D.~Dutta, G.~Huber and T.~Horn    
for making the actual $(W,Q^2)$ values of the data available to us. 
We also gratefully acknowledge helpful communications and discussions 
with H.~Blok, R.~Ent and H.~Gao.

This work was supported by BMBF.
\end{acknowledgments}

\end{document}